\documentclass[sigconf,authorversion,nonacm,screen]{acmart}

\AtBeginDocument{%
  \providecommand\BibTeX{{%
    \normalfont B\kern-0.5em{\scshape i\kern-0.25em b}\kern-0.8em\TeX}}}

\usepackage{soul}
\usepackage{balance}

\setcopyright{rightsretained}
\acmPrice{}
\acmDOI{10.1145/3573382.3616051}
\acmYear{2023}
\copyrightyear{2023}
\acmSubmissionID{playcomp23pop-p1013-p}
\acmISBN{979-8-4007-0029-3/23/10}
\acmConference[CHI PLAY '23]{Companion Proceedings of the Annual Symposium on Computer-Human Interaction in Play}{October 10--13, 2023}{Stratford, ON, Canada}
\acmBooktitle{Companion Proceedings of the Annual Symposium on Computer-Human Interaction in Play (CHI PLAY '23), October 10--13, 2023, Stratford, ON, Canada}
\received{2023-06-22}
\received[accepted]{2023-08-03}

\begin{document}

\title[Super Synthesis Pros.]{Super Synthesis Pros., or\\ Why CHI PLAY Needs Research Synthesis}

\author{Katie Seaborn}
\email{seaborn.k.aa@m.titech.ac.jp}
\orcid{0000-0002-7812-9096}
\affiliation{%
  \institution{Tokyo Institute of Technology}
  \city{Tokyo}
  \country{Japan}
}

\renewcommand{\shortauthors}{Seaborn}

\begin{abstract}
  Games user research is a-booming---or maybe a-goomba-ing---with a boundless parade of papers popping up from every nook and pipe. We may need a super power---or super method---from another world. I outline three motivations for jump-starting research synthesis in games user research. I argue that:
  research synthesis will validate this field of study and enrich primary research (meta-scholarship); 
  we must level up both primary and secondary research (education); and 
  we should reflect this epistemological stance in community structures and adopt established tools and protocols (standardization). I offer power-ups to get the toads rolling.
\end{abstract}




\received{20 February 2007}
\received[revised]{12 March 2009}
\received[accepted]{5 June 2009}

\maketitle

\section{Introduction to the Mushroom Synthdome}

\emph{Super Mario} is a transmedia phenomenon, with titular character Mario going on to star in movies, television series, comics, and cartoons. The appeal spans cultures and generations, offshoots and (re)incarnations, a plethora of merchandise, and even a theme park 
in Japan. Similarly, games user research is a transdisciplinary area of study: anchored to a wealth of philosophical bases, drawing from a variety of methodological traditions, and contributing to a range of knowledge sectors spanning computer science, psychology, and the humanities. 
Like its fountainhead, human-computer interaction (HCI), games user research is entering a phase characterized by a long-running conference, recognition within and beyond the field, and an increasing number of papers being published each year \cite{rogers2023systematic}. I argue that now is the time for a transformation in community positionality and practice. In short, \emph{games user research needs research synthesis, and pronto}. Like Mario, a simple plumber finding himself in a strange new world, we can approach this moment with curiosity and bravery, even if there may be growing pains and not a few missing princesses along the way.

Research synthesis, like the Mushroom Kingdom, treads ground both familiar and uncanny. An umbrella term, it refers to meta-methods that aim to capture and synthesize the literature on a particular topic \cite{krnic2019definition}. The purpose is to bring together primary research outcomes in a meaningful way and ideally in a format for communication within and beyond the field. Here, I motivate the uptake of research synthesis work within games user research. As a base of human knowledge, our field requires research synthesis and may benefit from a customized approach--its own "Koopa Troopa" of "survey" soldiers, heralding from other domains but acclimatized to this "level" of study. We must "hammer" this initiative into our "world" in a community-oriented and community-driven way. I sound the call for a power-up.


\section{Three Motivations to Power-Up Research Synthesis in Games User Research}

\subsection{Meta-Scholarship: Beware of Pixels and Pantomimes}
Mario's distinctive look---a mustachioed ball of red, white, and blue---has become iconic. Yet, it emerged from the era's technical shortcomings: limited pixels and colour sets, restrictions on animation. Even after technical advances and many iterations on the franchise, the creators have stayed the minimalist course,
adding novel aspects, such as the game-changing wall-jumping ability, with prudence. This formula of restraint is a time-tested success in the case of Mario. The impact of the series and character on gaming in general cannot be understated.

Novelty is a driver of much research. Calls for novelty and innovation pepper the author instructions at CHI PLAY and adjacent venues. But new ideas are not necessarily good for knowledge generation and advancing practice \cite{balestrini2015civically}. Publication biases \cite{hornbaek2015we}, replication crises \cite{shrout_psychology_2018}, p-hacking \cite{head_extent_2015}, and preprint mayhem \cite{noauthor_rise_2020} are disastrous side-effects, and games user research will not be exempt. At the same time, the status quo of what games user research "is" can become entrenched \cite{chu2021slowed}: Goomba after Goomba after Goomba, soaking up Mario's attention. We need to zoom out and take in the whole level. Like well-honed character design and level-crafting, research synthesis can guide the choice of studies, balance novelty against established tracks, and allow those outside of the field to grasp the fundamentals.
\vspace{12pt}

\noindent \includegraphics[height=\fontcharht\font`\B]{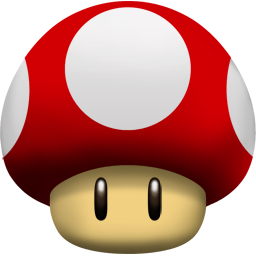} \textbf{Power-ups:}\begin{itemize}
\item {\emph{Time Limit}}: Games user research is at a critical juncture: a young but maturing field. We do not want to find ourselves in a "paperdemic" \cite{dinis2020covid}, unable to justify our worth to the larger research community. The timer is counting down, and we have many levels to traverse. We must take \includegraphics[height=\fontcharht\font`\B]{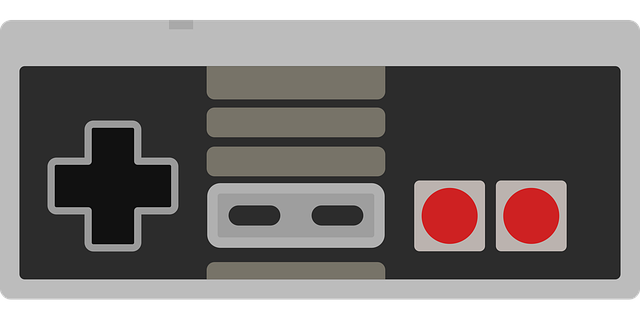} now.

\item {\emph{Stomp Novelty}}: Let us embrace novelty but not to the extent that we shift the focus away from better understanding known player phenomena and replicating results. Even Mario has to slow down at points, to creep over quicksand or crawl along rooftops. Slow science \cite{stengers_another_2018} has a place here.

\item {\emph{World Select Menu}}: We need a resource hub to encourage community-centrism and discourage reinventing the wheel. CHI PLAY can also offer a publication track similar to HRI's Short Contributions\footnote{\url{https://humanrobotinteraction.org/2022/short-contributions/}}.
\end{itemize}


\subsection{Education: Becoming Super Synthesis Pros.}
We have warped into a strange new world. But we need not be fearful. Like Mario, we have a suite of power-ups---methods, protocols, tools, and templates---to assist us in becoming super synthesis pro(fessional)s. For this, we need to consider two broad perspectives: \emph{primary} research and \emph{secondary} research synthesis. Some will wear both hats; others, one. Still, we must work in sync to enable knowledge-building. Even so, an informal survey of the HCI community (n=48) indicated that people desire more consistency and rigour in how research synthesis is done\footnote{Results @ HCI Systematic Review-lution: \url{https://ajkt.github.io/sysreviewlution/informalsurvey2023.html}}. Primary research is not merely coins to collect, nor is secondary research as forbidding as Bowser and his minions. We all participate in an ecology of research praxis, and as players, we owe it to each other to learn about and value our place in this world.
\vspace{12pt}

\noindent \includegraphics[height=\fontcharht\font`\B]{mush3dsm.png} \textbf{Power-ups:}\begin{itemize}
\item {\emph{Tutorial Level}}: We can world-hop between the initial guides offered in medicine and health (e.g., Tawfik et al. \cite{tawfik2019step} may be especially relevant, since it is focused on simulation data) and other media, such as YouTube\footnote{For example, the guides by Research Shorts (e.g., \url{https://youtu.be/WUErib-fXV0}) and Dr. Amina Yonis (e.g., \url{https://youtu.be/TLvF80WIXX8}).}.
\item {\emph{Sandbox}}: We can learn by doing: tracing the players before us by replicating existing syntheses (e.g., how "WEIRD" and "uncanny" \cite{seaborn2023not} is CHI PLAY research ...?) or playing with open data sets\footnote{To take random yet awesome example of curation: \url{https://github.com/leomaurodesenv/game-datasets}}, like the longitudinal PowerWash Simulator data set\footnote{\url{https://psyarxiv.com/kyn7g}} and the PXI (Player Experience Inventory) Benchmark Data\footnote{\url{https://playerexperienceinventory.org/bdata}}.
\end{itemize}


\subsection{Standardization: Synth-o-Ware, from Warp Pipes and Wing Caps to Hammers and Hurly Gloves}

Games user research, like HCI \cite{rogers2023systematic}, is characterized by unique "worlds."
We should adopt standardized tools--but also consider whether and how they need to be transformed. We hail from a variety of backgrounds and have a range of epistemological stances. Indeed, I chose "research synthesis" over other terms because this diversity should be embraced.
\vspace{12pt}

\noindent \includegraphics[height=\fontcharht\font`\B]{mush3dsm.png} \textbf{Power-ups:}
\begin{itemize}
\item {\emph{Phenomenon Registration}}: Phenomena of interest, subjects of study, hypotheses, and measured constructs should be registered at a central repository. Primary research should be linked to each registration on publication. Examples include validated measures, such as the PXI \cite{abeele2020development} and its "chibi" version \cite{haider2022minipxi}, as well as others frequently used in the field (refer to Brühlmann and Mekler \cite{bruhlmann2018surveys}), accounts of experiences with specific games (compare Tetris to League of Legends), and player behaviours, such as cheating \cite{passmore2020cheating} or (too afraid to cite) violence. 

\item {\emph{Reporting Structures:}} Data are meaningless without \emph{framing}. CHI PLAY advanced the notion of citing games, but these guidelines mysteriously disappeared after 2020\footnote{\url{https://chiplay.acm.org/2020/guidelines/}} and need refining, such as for series. Gerling and Birk \cite{gerling2022reflections} advocated for standardizing how games as research data are described. Metadata and schema may also be used for automated synthesis and need standards to be adapted for the particularities of games user research.

\item {\emph{Primary Reporting Tools}}: We can adapt reporting frameworks and schema, such as SPIDER \cite{methley2014pico} or ROSES \cite{haddaway2018roses}, decided in workshops to instill community norms and encourage their use by publishers in official templates.

\item {\emph{Secondary Reporting Tools}}: We need to identify a selection of reporting frameworks, such as PRISMA \cite{moher2009preferred}, and adapt these to games user research using a similar approach as for primary reporting tools.
\end{itemize}

\section{Conclusions and Next Stomps}
Mario often pops a Super Mushroom when starting a new adventure. I offer this set of \st{mushrooms} motivations for the community to munch on. Will we hit the \includegraphics[height=\fontcharht\font`\B]{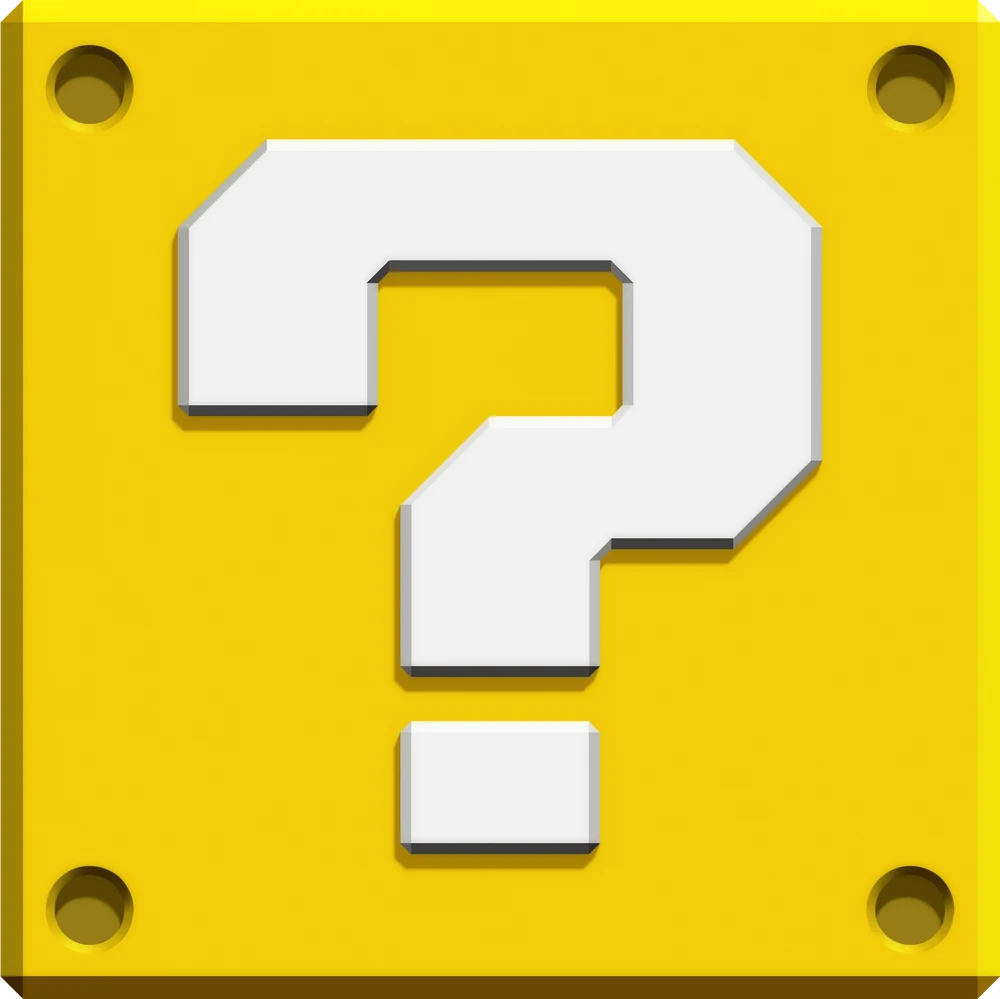} block and power-up? Or will we barrel headlong into a shell-full troop of unknown number? Research synthesis need not be as elusive as a princess in another castle.



\begin{acks}
Cappies off to fellow systematic review-lutionary, Dr. Katja Rogers, for inspiration and sundry. I also thank the chairs and the external peer reviewers for jumping on board and augmenting the decision-making process on this paper. Nintendo-esque icons were sourced from pngwing (non-commercial use) and fall under fair use.
\end{acks}


\bibliographystyle{ACM-Reference-Format}
\balance
\bibliography{main.bbl}









\end{document}